\documentstyle[12pt,epsf]{article}
\begin{document}

{}~ \hfill BIHEP-96-38

{}~ \hfill TECHNION-PH-96-25

{}~ \hfill (Revised)

\begin{center}
{\large\bf  W-exchange and W-annihilation processes of B mesons}\\[18mm]
{\bf Dongsheng Du$^{a,b}$, Li-bo Guo$^{b,c}$ and Da-Xin Zhang$^{d}$}\\[7mm]
$^a$ CCAST (World Laboratory),
         P.O.Box 8730, Beijing 100080, China\\[3mm]
$^b$ Institute of High Energy Physics, Chinese Academy of Sciences\\
P.O.Box  918(4),  Beijing, 100039, P. R. China\footnote{mailing address.}\\[3mm]
$^c$ Department of Physics, Henan Normal University\\
Xinxiang, 453002, Henan, P.R. China\\[3mm]
$^d$ Department of Physics,
Technion -- Israel Institute of Technology,
Haifa 32000, Israel\\[18mm]
{\bf Abstract}
\end{center}

\vspace{1cm}

Using the PQCD method
we calculate the W-exchange and the W-annihilation processes of B mesons,
which in general involve a charm quark or anti-quark in the final state.
The nonvanishing amplitudes of these processes are found to be suppressed
by a factor of $m_c/m_b$ compared to the tree or the time-like penguin
processes, but some of them are within the  reach of observation at 
the future B-factories,
and $\bar B_d^0 \to D^+_s K^-$ whose branching ratio is found to be 
$6.6 \times 10^{-6}$ can be found even before the B-factory era.
Comparisons with the results based on the BSW model are also given.

 \vspace{0.4cm}
\noindent {\it PACS number(s):}  13.20.Hw, 12.38.Bx, 12.15.Ji

\noindent {\it Keywords:} W-exchange processes, W-annihilation processes

\newpage

In the forthcoming years, 
more data of B decay
processes will be available at Tevatron, CLEO and 
at the B-factories.
Rare hadronic B decays dominated by the
W-exchange or the W-annihilation diagrams
with the branching ratios
of $10^{-7}$ or $10^{-8}$ are possible to be measured,
 bringing the necessity of estimating 
 these kinds of processes in advance.
In the past,
most of the  theoretical investigations on the
nonleptonic processes are based on the
BSW model\cite{bsw},
where the factorization hypothesis has been
used together with some phenomenological inputs
such as $a_1$ and $a_2$\cite{bsw,a1a2}.
While the usefulness of the factorization hypothesis
in these rare decays
is doubtful,
it is also not certain that these phenomenological inputs
are independent of the processes;
that is to say,
using the parameters $a_1$ and $a_2$ extracted from
one experiment to other processes cannot be
taken for granted.

In the present study, 
we focus on the rare B decay processes
driven by the W-exchange or the W-annihilation diagrams.
Some of these processes have been calculated recently
within the BSW model in \cite{xing} where
very small branching ratios have been claimed.
Here we want to use the perturbative QCD (PQCD)
method\cite{pqcd,car2,wyler} to re-analyse them and compare
the results given by these two methods.
We follow directly Ref. \cite{wyler}
which takes the PQCD method as a 
phenomenologically acceptable model.
For instance, 
reasonable agreements have been arrived in the
observed modes $B\to K^* \gamma$
and $B\to K J/\psi$\cite{car2}.
In the present study of the W-exchange and the W-annihilation
processes,
we leave the applicability of this method
as an open question for the moment,
and will return to this issue at the very end. 

We denote the processes at the hadronic level as 
$\bar B_\delta \left( b \bar \delta \right) \rightarrow
Y(\alpha \bar \gamma) + Z(\gamma \bar \beta)$ ($\gamma=u,d,s$).
Their momenta  are denoted by $P_B$, $P_Y$ and $P_Z$, 
and  masses by $M_B$, $M_Y$ and $M_Z$, respectively.
The  effective Hamiltonian for these processes is\cite{buras}
\begin{equation}
{{\cal H}_{eff}} =4 \frac{G_F}{\sqrt 2}V_{ckm}
\left(C_1'(\mu) O_1 +C_2'(\mu) O_2\right),
\end{equation}
where the $CKM$ factor is defined as
\begin{eqnarray}
V_{ckm} &=& \left\{\begin{array}{ll}
            V_{\delta b}V^*_{\beta \alpha}  
                  & \mbox {for $W$-annihilation process,}\\
            V_{\alpha b}V^*_{\beta \delta}  
                  & \mbox {for $W$-exchange process,}
            \end{array}
            \right.               
\end{eqnarray}
the effective four-quark operators are
\begin{eqnarray}
O_1 &=& (\bar \delta \gamma_ \mu P_L \beta)
        (\bar \alpha \gamma_ \mu P_L b),\nonumber  \\
O_2 &=& (\bar \alpha \gamma_ \mu P_L \beta)
        (\bar \delta \gamma_ \mu P_L b),
\end{eqnarray}
$P_L=(1-\gamma_5)/2$.
 $\alpha, \beta=u ~{\rm or}~ c$, $\delta=d ~{\rm or}~ s$ for
the W-exchange processes, and
$\beta, \delta=u ~{\rm or}~ c$, $\alpha=d ~{\rm or}~ s$ for
the W-annihilation processes,
  and the Wilson coefficients are
\begin{eqnarray}
C_1'=C_1&,& C_2' = C_2  ~~{\rm for ~~W-exchange ~~process,}
\nonumber  \\
C_1' =C_2&,& C_2' = C_1  ~~{\rm for ~~W-annihilation ~~process.}
\end{eqnarray}
Calculation the Wilson coefficients at $\mu=5$GeV gives
$C_1=1.10,~~~C_2=-0.235$.

Following the PQCD method\cite{wyler},
 the figures which are relevant to the
 W-exchange or W-annihilation processes are depicted
in Fig 1.
We take the interpolating fields in the standard ways as\cite{pqcd,car2,wyler}
\begin{equation}
\psi_B=\displaystyle\frac{1}{\sqrt{2}}\displaystyle\frac{I_c}{\sqrt{3}}
       \phi_B(x)\gamma_5({\not\! P_B}-M_B),\nonumber
\end{equation}
\begin{equation}
\psi_{Y}=\displaystyle\frac{1}{\sqrt{2}}\displaystyle\frac{I_c}
        {\sqrt{3}}\phi_{Y}(y)\gamma_5 ({\not\!P_Y}+M_Y),
\end{equation}
\begin{equation}
\psi_{Z}=\displaystyle\frac{1}{\sqrt{2}}\displaystyle\frac{I_c}
        {\sqrt{3}}\phi_{Z}(z)\gamma_5 ({\not\!P_Z}+M_Z).\nonumber
\end{equation}
Here $I_c$ is  an identity in the color space.

The momentum fractions carried by every quark lines
are labeled in Fig 1 in an obvious way.
The  momentum flows for the gluon and for the quark
lines are
\begin{equation}
l_g = P_Y y_1 + P_Z z_2, 
q_b =P_B x_2 - l_g, 
q_\delta = -( P_B x_1 - l_g),  
q_\alpha = P_Y y_2 + l_g, 
q_\beta = -( P_Z z_1 + l_g).
\end{equation}
To present the results in a concise way, we also denote
\begin{equation}
\begin{array}{rclcl}
D_g &\equiv& l_g^2 
  &=& M_Y^2 y_1^2 + M_Z^2 z_2^2 + (M_B^2-M_Y^2-M_Z^2) y_1 z_2 ,\nonumber\\
D_b &\equiv& q_b^2 - m_b^2 
  &=& M_Y^2 y_1^2 +M_Z^2 z_2^2 
      -(M_B^2+M_Y^2-M_Z^2) x_2 y_1\nonumber\\ 
 & &  & &-(M_B^2-M_Y^2+M_Z^2) x_2 z_2 
   +(M_B^2-M_Y^2-M_Z^2) y_1 z_2 ,\nonumber\\
D_\delta &\equiv& q_\delta^2 - m_\delta^2
  &=& M_Y^2 y_1^2 +M_Z^2 z_2^2 +M_B^2 x_1^2 
      -(M_B^2+M_Y^2-M_Z^2) x_1 y_1   \nonumber \\
& &   & &    -(M_B^2-M_Y^2+M_Z^2) x_1 z_2 
      +(M_B^2-M_Y^2-M_Z^2) y_1 z_2 - m_\delta^2 ,\nonumber\\
D_\alpha &\equiv& q_\alpha^2 - m_\alpha^2
 &=& M_Y^2 +M_Z^2 z_2^2 
      +(M_B^2-M_Y^2-M_Z^2) z_2 - m_\alpha^2 ,\nonumber\\
D_\beta &\equiv& q_\beta^2 - m_\beta^2
 &=& M_Z^2 +M_Y^2 y_1^2 
      +(M_B^2-M_Y^2-M_Z^2) y_1 - m_\beta^2.
\end{array}
\end{equation}

Now we are in the position of calculating the decay amplitudes.
First,  we take the W-exchange diagram as an example.
 The decay amplitudes are calculated from Fig 1 (a) to Fig 1 (d). 
With the insertion of the operator  
$O_1 = (\bar \delta \gamma_ \mu P_L \beta)
        (\bar \alpha \gamma_ \mu P_L b) $,
the calculations of the four diagrams give:
\begin{eqnarray}
{\cal A}_a^1&=&\int_{0}^{1} [{\rm d}x][{\rm d}y][{\rm d}z]
     {\rm Tr}\left[  \displaystyle \psi_B \gamma_\mu P_L
     \psi_Z \left(i\frac{\lambda^a}{2}\gamma_\alpha g_s \right)
     \psi_Y \gamma^\mu P_L ({\not\! q_b}+m_b)
     \left(i\frac{\lambda^a}{2}\gamma^\alpha g_s \right) \right]
     \displaystyle\frac{-i}{D_b D_g}, \nonumber\\
{\cal A}_b^1&=&\int_{0}^{1} [{\rm d}x][{\rm d}y][{\rm d}z]
     {\rm Tr}\left[  \displaystyle \psi_B 
     \left(i\frac{\lambda^a}{2}\gamma_\alpha g_s \right)
     ({\not\! q_\delta}+m_\delta)\gamma_\mu P_L
     \psi_Z \left(i\frac{\lambda^a}{2}\gamma^\alpha g_s \right)
     \psi_Y \gamma^\mu P_L \right]
     \displaystyle\frac{-i}{D_\delta D_g}, \nonumber\\
{\cal A}_c^1&=&\int_{0}^{1} [{\rm d}x][{\rm d}y][{\rm d}z]
     {\rm Tr}\left[  \displaystyle \psi_B \gamma_\mu P_L
     \psi_Z \left(i\frac{\lambda^a}{2}\gamma_\alpha g_s \right)
     \psi_Y \left(i\frac{\lambda^a}{2}\gamma^\alpha g_s \right)
     ({\not\! q_\alpha}+m_\alpha)\gamma^\mu P_L \right]
     \displaystyle\frac{-i}{D_\alpha D_g}, \\
{\cal A}_d^1&=&\int_{0}^{1} [{\rm d}x][{\rm d}y][{\rm d}z]
     {\rm Tr}\left[  \displaystyle \psi_B \gamma_\mu P_L
     ({\not\! q_\beta}+m_\beta)
     \left(i\frac{\lambda^a}{2}\gamma_\alpha g_s \right)
     \psi_Z \left(i\frac{\lambda^a}{2}\gamma^\alpha g_s \right)
     \psi_Y \gamma^\mu P_L \right]
     \displaystyle\frac{-i}{D_\beta D_g}, \nonumber
\end{eqnarray}
where $[ {\rm d}x]$, $[ {\rm d}y]$, and $[ {\rm d}z]$ denote 
$({\rm d}x_1{\rm d}x_2)$, $({\rm d}y_1{\rm d}y_2)$ and
$({\rm d}z_1{\rm d}z_2)$, respectively.
With the insertion of  the operator 
$O_2 = (\bar \alpha \gamma_ \mu P_L \beta)
        (\bar \delta \gamma_ \mu P_L b)$, 
the results are:
\begin{eqnarray}
{\cal A}_a^2&=&\int_{0}^{1} [{\rm d}x][{\rm d}y][{\rm d}z]
   {\rm Tr}\left[
       \displaystyle\psi_B \gamma_\mu P_L (\not\! q_b+m_b)
       \left(i\displaystyle\frac{\lambda^a}{2}\gamma_\alpha g_s \right)\right]
   {\rm Tr}\left[  \displaystyle \psi_Z 
       \left(i\frac{\lambda^a}{2}\gamma^\alpha g_s \right)
       \psi_Y \gamma^\mu P_L \right]
    \frac{i}{D_b D_g},\nonumber\\
{\cal A}_b^2&=&\int_{0}^{1} [{\rm d}x][{\rm d}y][{\rm d}z]
   {\rm Tr}\left[  \displaystyle\psi_B  
         \left(i\displaystyle\frac{\lambda^a}{2}\gamma_\alpha g_s \right)
         (\not\! q_\delta+m_\delta)\gamma_\mu P_L \right]
   {\rm Tr}\left[  \displaystyle \psi_Z 
       \left(i\frac{\lambda^a}{2}\gamma^\alpha g_s \right)
       \psi_Y \gamma^\mu P_L \right]
    \frac{i}{D_\delta D_g},\nonumber\\
{\cal A}_c^2&=&\int_{0}^{1} [{\rm d}x][{\rm d}y][{\rm d}z]
   {\rm Tr}\left[  \displaystyle\psi_B \gamma_\mu P_L \right]
   {\rm Tr}\left[  \displaystyle \psi_Z 
           \left(i\displaystyle\frac{\lambda^a}{2}\gamma_\alpha g_s \right)
           \psi_Y \left(i\frac{\lambda^a}{2}\gamma^\alpha g_s \right)
           (\not\! q_\alpha+m_\alpha)\gamma^\mu P_L \right]
    \frac{i}{D_\alpha D_g},\\
{\cal A}_d^2&=&\int_{0}^{1} [{\rm d}x][{\rm d}y][{\rm d}z]
   {\rm Tr}\left[  \displaystyle\psi_B \gamma_\mu P_L \right]
   {\rm Tr}\left[  \displaystyle \psi_Z 
           \left(i\displaystyle\frac{\lambda^a}{2}\gamma_\alpha g_s \right)
           \psi_Y \gamma^\mu P_L (\not\! q_\beta+m_\beta) 
           \left(i\frac{\lambda^a}{2}\gamma^\alpha g_s \right)\right]
    \frac{i}{D_\beta D_g},\nonumber
\end{eqnarray}

Performing the trace operation in both the spinor and the color space, 
we find that the contributions of ${\cal A}_a^2$, ${\cal A}_b^2$ vanish
due to their color  structures,
thus
\begin{eqnarray}
{\cal A}_a^1&=&-\frac{8}{3 \sqrt 6} g_s^2
  \int_{0}^{1} [{\rm d}x][{\rm d}y][{\rm d}z] \Psi(x,y,z) 
  \left[P_{BY}(P_{BZ} x_2 - P_{YZ} y_1 - 2 M_Z^2 z_2)- 
  M_B^2 P_{YZ} x_2\right] \frac{1}{D_b D_g}, \nonumber \\
{\cal A}_b^1&=&-\frac{8}{3 \sqrt 6} g_s^2
  \int_{0}^{1} [{\rm d}x][{\rm d}y][{\rm d}z] \Psi(x,y,z) 
  \left[P_{BZ}(P_{YZ} z_2 - P_{BY} x_1 + 2 M_Y^2 y_1)+ 
  M_B m_\delta P_{YZ}\right] \frac{1}{D_\delta D_g}, \nonumber \\
{\cal A}_c^1&=&-\frac{8}{3 \sqrt 6} g_s^2
  \int_{0}^{1} [{\rm d}x][{\rm d}y][{\rm d}z] \Psi(x,y,z)
   \left[P_{BY} M_Z^2 z_2-P_{BZ}(M_Y^2+P_{YZ} z_2)\right. \nonumber \\
&& \left.  + m_\alpha (M_Z P_{BY}+ 2 M_Y P_{BZ}) \right] 
   \frac{1}{D_\alpha D_g}, \\
{\cal A}_d^1&=&-\frac{8}{3 \sqrt 6} g_s^2
  \int_{0}^{1} [{\rm d}x][{\rm d}y][{\rm d}z] \Psi(x,y,z)
   \left[P_{BY}(P_{YZ} y_1 +  M_Z^2) -P_{BZ} M_Y^2 y_1 \right. \nonumber \\
&& \left. - m_\beta (2 M_Z P_{BY}+ M_Y P_{BZ}) \right] 
\frac{1}{D_\beta D_g}, \nonumber \\
{\cal A}_a^2&=&0, ~~
{\cal A}_b^2~=~0, ~~
{\cal A}_c^2= 3{\cal A}_c^1, ~~{\cal A}_d^2~~=~~3 {\cal A}_d^1,\nonumber
\end{eqnarray}
where  $\Psi(x,y,z)~=~\phi_B(x) \phi_Y(y) \phi_Z(z)$ and
$P_{ij}~=~2 P_i \cdot P_j$. 
We have set $M_Y M_Z~=~0$ 
since there are  light mesons in the final states.

Using the effective Hamiltonian (1) for
$\bar B_\delta \left( b \bar \delta \right) \rightarrow 
Y(\alpha \bar \gamma) + Z(\gamma \bar \beta)$ 
decays, then the decay amplitude can be written as
\begin{equation}
{\cal A}~=~4 \frac{G_F}{\sqrt 2} V_{ckm}
\left[ C_1'({\cal A}_a^1+{\cal A}_b^1)+(C_1' +3 C_2')({\cal A}_c^1 +
{\cal A}_d^1)\right]
\end{equation}

The W-exchange processes can be divided into two cases: 
one is $\alpha~=~u$ and $\beta~=~c$, 
the other is $\alpha~=~c$ and $\beta~=~u$.
Three  processes belonging to the former,
they are $\bar B_d^0 \rightarrow D_s^-K^+$, $\bar B_s^0 
\rightarrow D^-\pi^+$ and $\bar B_s^0 \rightarrow \bar D^0 \pi^0$.
In this case, because the $Y$ meson is a light meson, 
we set $M_Y~=~0$. 
The decay amplitudes can be simplified as
\begin{eqnarray}
{\cal A}_a^1&=&-\frac{8}{3 \sqrt 6} g_s^2 
  \int_{0}^{1} [{\rm d}x][{\rm d}y][{\rm d}z] \Psi(x,y,z) 
\frac{1}{z_2}
\frac{y_1-\frac{\Delta_Z (x_2-2 z_2)}{1-\Delta_Z}}
{(y_1+\frac{\Delta_Z z_2}{1-\Delta_Z})\left[(x_2-z_2)y_1+
\frac{(1+\Delta_Z)x_2 z_2-\Delta_Z z_2^2}{1-\Delta_Z} \right]}, \nonumber \\
{\cal A}_b^1&=&-\frac{8}{3 \sqrt 6} g_s^2 
  \int_{0}^{1} [{\rm d}x][{\rm d}y][{\rm d}z] \Psi(x,y,z) 
\frac{1}{z_2}\frac{1}{(y_1+\frac{\Delta_Z z_2}{1-\Delta_Z})
(y_1-\frac{x_1-\Delta_Z z_2}{1-\Delta_Z})}
\frac{1+\Delta_Z}{1-\Delta_Z}, \nonumber \\
{\cal A}_c^1&=&-\frac{8}{3 \sqrt 6} g_s^2 
  \int_{0}^{1} [{\rm d}x][{\rm d}y][{\rm d}z] \Psi(x,y,z) 
\frac{1}{z_2}
  \frac{-1}{(1-\Delta_Z + \Delta_Z z_2)
(y_1+\frac{\Delta_Z z_2}{1-\Delta_Z})}, \\
{\cal A}_d^1&=&-\frac{8}{3 \sqrt 6} g_s^2
  \int_{0}^{1} [{\rm d}x][{\rm d}y][{\rm d}z] \Psi(x,y,z) 
\frac{1}{  z_2}
\frac{y_1-\frac{\Delta_Z (1- 2 z_2)}{1-\Delta_Z}}
{(y_1+\frac{\Delta_Z z_2}{1-\Delta_Z})
(y_1+\frac{\Delta_Z z_2 (2-z_2)}{1-\Delta_Z})},\nonumber
\end{eqnarray}
where $\Delta_Z=m_Z^2/M_B^2$.

The later case of $\alpha~=~c$ and $\beta~=~u$ also has three
corresponding processes which are
 $\bar B_d^0 \rightarrow D_s^+K^-$, $\bar B_s^0 
\rightarrow D^+\pi^-$ and $\bar B_s^0 \rightarrow  D^0 \pi^0$.
Again the light meson mass $M_Z~=~0$,
and the decay amplitudes are simplified into
\begin{eqnarray}
{\cal A}_a^1&=&-\frac{8}{3 \sqrt 6} g_s^2 
  \int_{0}^{1} [{\rm d}x][{\rm d}y][{\rm d}z] \Psi(x,y,z) 
 \frac{1}{y_1}\frac{y_1-\frac{\Delta_Y x_2}{1+\Delta_Y}}
 {(x_2-y_1)z_2+\frac{x_2y_1+\Delta_Y y_1(x_2-y_1)}{1-\Delta_Y}}
 \frac{1+\Delta_Y}{1-\Delta_Y},\nonumber\\
 {\cal A}_b^1&=&-\frac{8}{3 \sqrt 6} g_s^2 
  \int_{0}^{1} [{\rm d}x][{\rm d}y][{\rm d}z] \Psi(x,y,z) 
   \frac{1}{y_1}\frac{z_2-\frac{x_1+\Delta_Y(x_1-2 y_1)}{1-\Delta_Y}}
   {(y_1-x_1)(z_2+\frac{\Delta_Y y_1}{1-\Delta_Y})
   (z_2-\frac{x_1-\Delta_Y y_1}{1-\Delta_Y})},\nonumber \\
{\cal A}_c^1&=&-\frac{8}{3 \sqrt 6} g_s^2 
  \int_{0}^{1} [{\rm d}x][{\rm d}y][{\rm d}z] \Psi(x,y,z) 
 \frac{1}{y_1}\frac{-(z_2-\frac{(1-2 y_1)\Delta_Y}{1-\Delta_Y})}
 {(z_2+\frac{\Delta_Y y_1}{1-\Delta_Y})
 (z_2+\frac{y_1(2-y_1)\Delta_Y}{1-\Delta_Y})},\\
 {\cal A}_d^1&=&-\frac{8}{3 \sqrt 6} g_s^2
  \int_{0}^{1} [{\rm d}x][{\rm d}y][{\rm d}z] \Psi(x,y,z) 
 \frac{1}{y_1}\frac{1}{(1-\Delta_Y+y_1\Delta_Y)
 (z_2+\frac{\Delta_Y y_1}{1-\Delta_Y})},\nonumber
\end{eqnarray}
where $\Delta_Y=M_Y^2/M_B^2$.

Next, we turn to the pure W-annihilation processes
which differ from the W-exchange processes in the $CKM$ factor and 
the Wilson coefficients as described in Eq.~(2) and Eq.~(4),respectively.
There are also two cases for the pure annihilation processes:
one is   $\delta~=~u$  and the other is $\delta~=~c$,
corresponding to the W-annihilation processes of  $B_u^-$ into a $\bar D$
meson plus a light one,
and of $B_c^-$ into two light mesons, respectively.
The analyses of the later case of $B_c^-$ processes need more likely
to be performed within the potential model\cite{chang}, 
which is beyond the present investigation.
In the former case, we have
\begin{eqnarray}
{\cal A}_a^1&=&-\frac{8}{3 \sqrt 6} g_s^2 
  \int_{0}^{1} [{\rm d}x][{\rm d}y][{\rm d}z] \Psi(x,y,z) 
\frac{1}{z_2}
\frac{y_1-\frac{\Delta_Z (x_2-2 z_2)}{1-\Delta_Z}}
{(y_1+\frac{\Delta_Z z_2}{1-\Delta_Z})\left[(x_2-z_2)y_1+
\frac{(1+\Delta_Z)x_2 z_2-\Delta_Z z_2^2}{1-\Delta_Z} \right]}, \nonumber \\
{\cal A}_b^1&=&-\frac{8}{3 \sqrt 6} g_s^2 
  \int_{0}^{1} [{\rm d}x][{\rm d}y][{\rm d}z] \Psi(x,y,z) 
\frac{1}{z_2}\frac{1}{(y_1+\frac{\Delta_Z z_2}{1-\Delta_Z})
(y_1-\frac{x_1-\Delta_Z z_2}{1-\Delta_Z})}
\frac{1+\Delta_Z}{1-\Delta_Z}, \nonumber \\
{\cal A}_c^1&=&-\frac{8}{3 \sqrt 6} g_s^2 
  \int_{0}^{1} [{\rm d}x][{\rm d}y][{\rm d}z] \Psi(x,y,z) 
\frac{1}{z_2}
  \frac{-1}{(1-\Delta_Z + \Delta_Z z_2)
(y_1+\frac{\Delta_Z z_2}{1-\Delta_Z})}, \\
{\cal A}_d^1&=&-\frac{8}{3 \sqrt 6} g_s^2
  \int_{0}^{1} [{\rm d}x][{\rm d}y][{\rm d}z] \Psi(x,y,z) 
\frac{1}{  z_2}
\frac{y_1-\frac{\Delta_Z (1- 2 z_2)}{1-\Delta_Z}}
{(y_1+\frac{\Delta_Z z_2}{1-\Delta_Z})
(y_1+\frac{\Delta_Z z_2 (2-z_2)}{1-\Delta_Z})},\nonumber
\end{eqnarray}
%

To get the numerical estimations,
we choose the wavefunctions for the mesons as
\begin{eqnarray}
\phi_B(x)&=&\frac{f_B}{2{\sqrt 3}}\delta(x-\epsilon_B),\nonumber\\
\phi_D(x)&=&\frac{f_D}{2{\sqrt 3}}\delta(x-\epsilon_D),\nonumber\\
\phi_K(x)&=&{\sqrt 3}f_K x(1-x),\nonumber\\
\phi_{\pi}(x)&=&{\sqrt 3}f_{\pi} x(1-x).
\end{eqnarray}
We will take in the numerical calculations
$\epsilon_B=0.07$,  $\epsilon_{B_s}=0.09$,
$\epsilon_D=0.2$ and $\epsilon_{D_s}=0.25$
Under these choices,
all the nonvanishing decay amplitudes  considered here
are proportional to $1/\epsilon_{D_{(s)}}$.
This observation is a common feature of the
peak approximations of the wavefunctions for the heavy mesons,
independent of the choices of the wavefunctions for the light mesons.
Note that from the previous studies,
the amplitudes for the tree diagrams and for the time-like penguins
are proportional to $1/\epsilon_B$\cite{wyler},
while there is no such enhancement in the amplitudes for the 
space-like penguins\cite{lu}.
Here what we find is that the amplitudes for the pure W-exchange 
or the W-annihilation processes,
which in general involve  $D$ or $\bar D$ mesons in the final states,
are suppressed by a factor of $\epsilon_B/\epsilon_D\sim m_c/m_b\sim 1/3$
compared to the tree or time-like penguin amplitudes.
The reason for these results can be understood using
the same arguments which induce  the helicity suppression mechanism\cite{heli},
as has been supposed in the quark diagram scheme in \cite{chau}.

To compare with the results given by \cite{xing},
we adopt the same numerical values for the
CKM matrix elements, for the mass parameters and
for the 
decay constants\footnote{Note that 
our definitions of the decay constants differ from those
in \cite{xing} by a factor of $1/\sqrt{2}$.} 
as those used in \cite{xing}.
Our results are presented in Table 1.

It can be observed from Table 1 that,
except the double CKM suppressed process $B^-  \to D_s^- K^0$,
all  these W-exchange and W-annihilation processes are within
the reach of discovery at
the future B-factories\footnote{
At the designed SLAC B-factory,
$3\times 10^{8}$ pairs of $B_s\bar B_s$ and more
non-strange B meson pairs can be produced\cite{slac}.},
since our results are in general larger than the predictions
made within the BSW model\cite{xing}.
Among them, the process $\bar B_d^0 \to D^+_s K^-$ whose
branching ratio  is found here to be $ 6.6 \times 10^{-6}$,
might be  discovered even before the B-factory era
based on our predictions.

Now we discuss the applicability of the PQCD method
in the W-exchange and W-annihilation processes we have focused on.
In \cite{wyler}  lower cuts on the integrations over the
momentum fractions   have been enforced to avoid
all possible poles; 
consequently, the numbers given in \cite{wyler} 
 are of orders smaller than those given in the BSW model.
Different from \cite{wyler},
in \cite{car2} no cut has been required and some numbers are
compareble with the data.
In our  case,
there exists only the $\bar \delta$ pole (see Figure 1(b));
all other poles, especially the gluonic pole,
lie out of the  range [0,1] of the integration.
A safe lower cut to avoid this $\bar \delta$ pole  in all these channels 
is  0.1 for the momentum fraction of the light meson.
We also give   our  predictions in Table 1
based on this cut.
Comparing to those without the cut,
it can be seen that the branching ratios are reduced
by $20\%$ to $60\%$.
These changes are quite moderate to justify the method, 
if  comparisions with the processes considered in \cite{wyler}
are made.

\vskip 0.6cm
We thank M. Gronau for  discussion which enlightened the present work.
This research of DXZ was supported in part by Grant 5421-3-96
from the Ministry of Science and the Arts of Israel.

\newpage
\begin{table}
\begin{tabular}{l|l|ll}
\hline
\hline
Process & Ref. \cite{xing} & This work$^a$&This work$^b$\\
\hline
$B^-  \to D^-\bar K^0$ &  $8.1\times 10^{-9}$ &  
$1.0 \times 10^{-8}$ &  $4.9 \times 10^{-9}$\\
$B^-  \to D_s^- K^0$   &  $4.2\times 10^{-10}$  &  
$5.5 \times 10^{-10}$ &  $5.1 \times 10^{-10}$\\
$\bar B_d^0 \to D^+_s K^-$ &$6.5\times 10^{-8}$  &   
$ 6.6 \times 10^{-6}$   &   $ 4.7 \times 10^{-6}$\\
$\bar B_d^0 \to D_s^- K^+$ &$2.1\times 10^{-11}$  &  
$3.5 \times 10^{-9}$   &  $1.2 \times 10^{-9}$\\
$\bar B_s^0 \to D^+\pi^-$  &$1.2\times 10^{-8}$  &  
$5.8 \times 10^{-7}$   &  $4.7 \times 10^{-7}$\\
$\bar B_s^0 \to D^0  \pi^0$ &$1.2\times 10^{-8}$  &  
$ 2.9 \times 10^{-7}$  &  $2.3 \times 10^{-7}$\\
$\bar B_s^0 \to D^- \pi^+$  &$1.5\times 10^{-9}$ &  
$ 4.9 \times 10^{-8}$   &  $ 3.2 \times 10^{-8}$\\
$\bar B_s^0 \to \bar D^0 \pi^0$ & $1.5\times 10^{-9}$&  
 $ 2.5 \times 10^{-8}$ &   $ 1.6 \times 10^{-8}$\\
\hline
\hline
\end{tabular}
\end{table}

\noindent
{\bf  Table 1.} 
 Comparison of the results with those of \cite{xing}. \\
$a$: no cut enforced;
$b$: lower cut at 0.1.

\vskip 0.6cm
\noindent
{\bf  Figure 1.} Diagrams which are relevant in PQCD calculations of
the W-exchange and the W-annihilation processes.
The solid blob denotes an insersion of the four-quark operator
$O_1$ or $O_2$.
Momentum fractions are labelled as $x$, $1-x$, etc.

\begin{figure}[htb]
\centerline{\epsfxsize6in\epsfbox{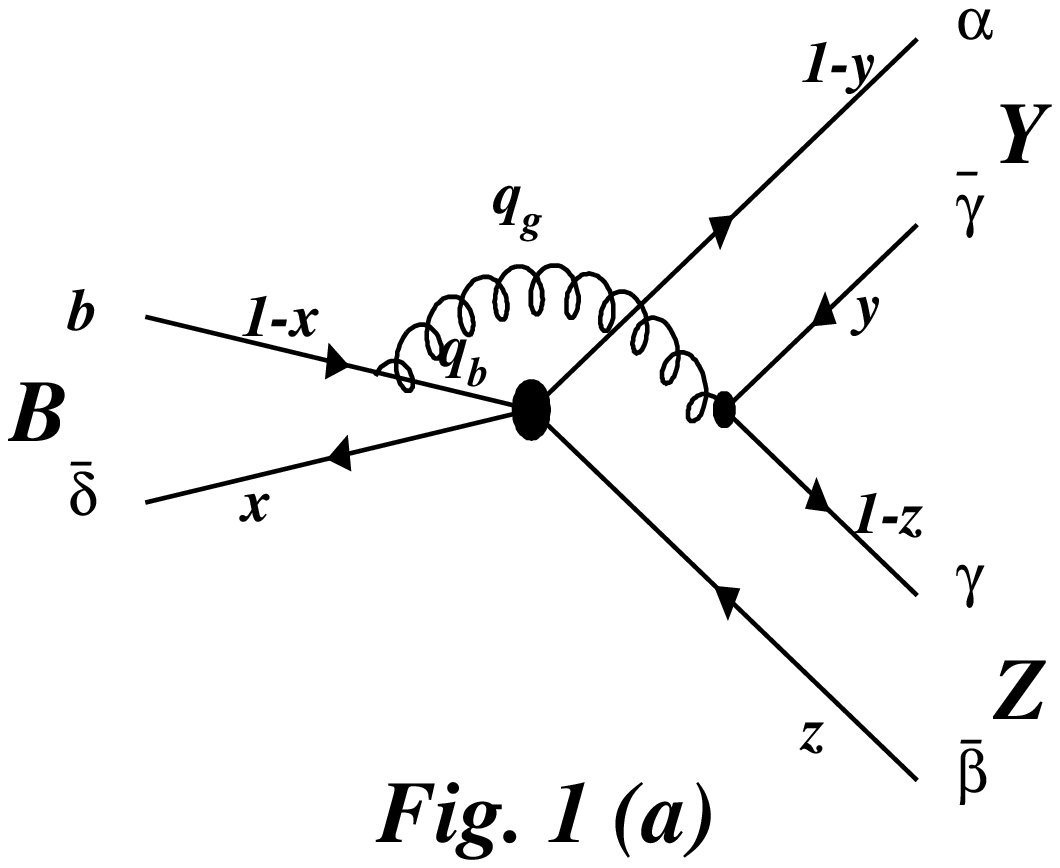}}
\vskip -5in
\centerline{\epsfxsize6in\epsfbox{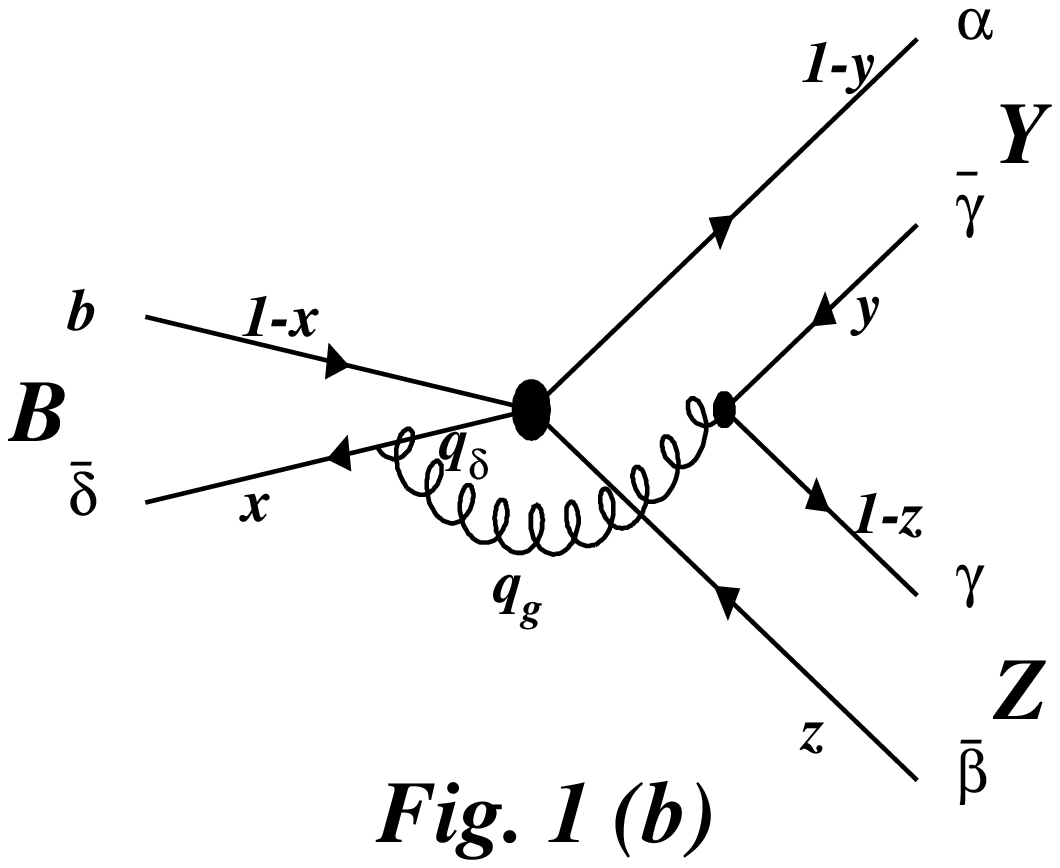}}
\end{figure}

\begin{figure}[htb]
\centerline{\epsfxsize6in\epsfbox{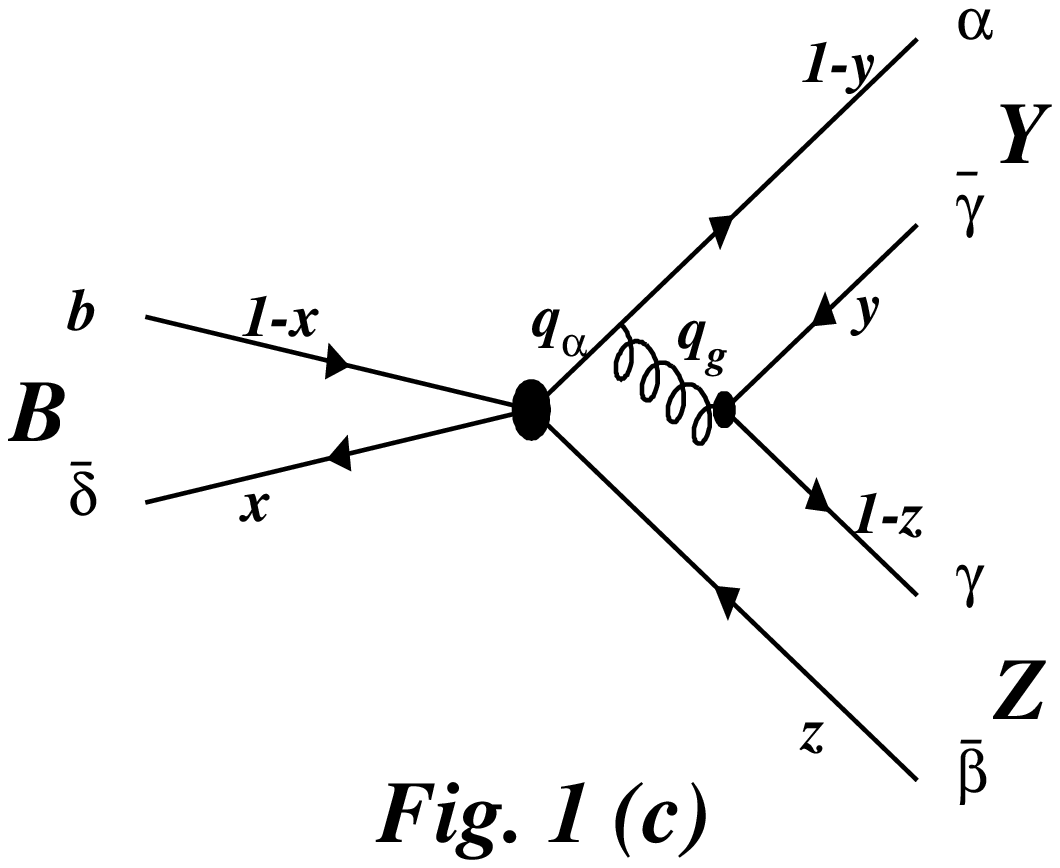}}
\vskip -5in
\centerline{\epsfxsize6in\epsfbox{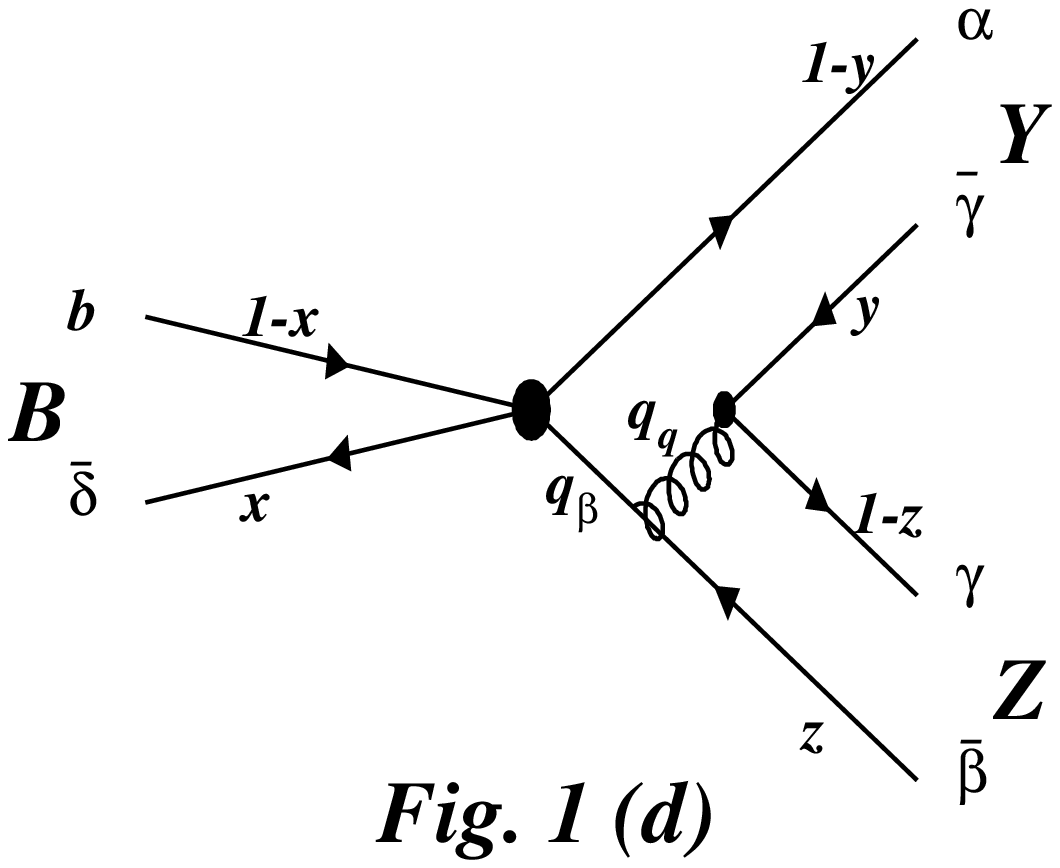}}
\end{figure}

\end{document}